\documentclass[journal]{IEEEtran}
\IEEEoverridecommandlockouts

\usepackage[cmex10]{amsmath}

\usepackage{enumitem}
\usepackage{graphicx}
\usepackage{color}
\usepackage{amsmath}
\usepackage{mathtools}
\usepackage{multicol}
\usepackage{multirow}
\usepackage[english]{babel}
\usepackage{blindtext}
\usepackage{algorithm}
\usepackage{algorithmic}
\usepackage{balance}
\usepackage{amsfonts}
\usepackage{bm}
\usepackage{stfloats}
\usepackage{subfig}
\usepackage{amsthm}
\usepackage{amssymb}
\usepackage{setspace}
\usepackage[nosort]{cite}
\usepackage{CJK}
\usepackage{cite}
\usepackage{caption}
\usepackage{comment}
\usepackage{array,multirow}
\usepackage{graphicx}

\theoremstyle{plain}

\usepackage[table]{xcolor}

\begin{document}

\captionsetup[figure]{labelformat={default},labelsep=period,name={Fig.}}

\title{Satellite NOMA for Direct-to-Cell Communications: Fundamentals, Protocols, and Opportunities}

\author{Xiangyu Li, 
        Bodong Shang, 
        Junchao Ma, 
        Yuzheng Ren,
        Haijun Zhang
        and Pingyi Fan
\thanks{Xiangyu Li and Bodong Shang (corresponding author) are with the College of Information Science and Technology, Eastern Institute of Technology, Ningbo, Zhejiang 315200, China, and also with the State Key Laboratory of Integrated Services Networks, Xidian University, Xi’an 710071, China.}
\thanks{Junchao Ma is with the School of Electrical and Information Engineering, Jiangsu University of Technology, Changzhou 213001, China.} 
\thanks{Yuzheng Ren and Haijun Zhang are with the School of Computer and Communication Engineering, University of Science and Technology Beijing, Beijing 100083, China.} 
\thanks{Pingyi Fan is with the Department of Electronic Engineering, Beijing National Research Center for Information Science and Technology, Tsinghua University, Beijing 100084, China.}
}

\maketitle

\begin{abstract}
Direct-to-cell (DTC) satellite communication is regarded as one of the most recent technologies that provides global connectivity.
However, with the growing number of wireless users and devices, the design of DTC communications must satisfy the requirements of high-scale capabilities and efficient spectrum utilization.
To this end, integrating satellite communications with advanced multiple-access techniques, such as non-orthogonal multiple access (NOMA), has attracted considerable interest in developing NOMA-DTC communications.
In this article, we first introduce the fundamentals of NOMA-DTC communications, including architectural fundamentals, system design aspects, and potential applications.
Given the various cooperative modes and the still-evolving satellite network (SatNet) architectures, such as cooperative SatNets and multi-tier SatNets, we explore protocols that suit future SatNets and enhance system performance.
Furthermore, a case study is conducted to investigate the benefits of NOMA schemes for DTC communications and to compare them with OMA schemes.
Finally, to inspire further research, several opportunities for NOMA-DTC communications are presented.
\end{abstract}

\IEEEpeerreviewmaketitle

\section{Introduction}
Industry and academic researchers are seeking new network architectures to meet the increasing demand for ubiquitous and high-quality connectivity, particularly in remote and underserved regions. 
Considering the growing number of wireless users and devices in recent years, terrestrial networks (TNs) may face limitations, including high deployment costs, limited coverage, and poor scalability. 
TNs might be able to provide adequate services in urban areas; however, when they become limited or non-existent, they may fail to deliver reliable regional services in rural, maritime, or disaster-prone areas.
These limitations of TNs have led to extensive exploration of non-terrestrial networks (NTNs), especially satellite-based solutions.
In recent years, the development of direct-to-cell (DTC) satellite communications has shown great promise. DTC communications enable users to communicate directly with satellites, eliminating the need for ground stations and providing seamless global connectivity \cite{bakhsh2024multi}.

The rise of DTC communications is primarily attributed to the rapid development of low-Earth orbit (LEO) satellite constellations, such as Starlink, OneWeb, Kuiper, and Telesat. Compared with geostationary orbit (GEO) and medium-earth orbit (MEO) satellites, LEO satellites can be grouped to provide higher coverage and data rates with lower latency and lower production and launch costs.
Due to their lower altitude and faster orbital periods, LEO satellites enable more continuous connectivity and more frequent handovers. These characteristics make them highly suitable for direct communication with mobile users. 
Specifically, unlike traditional satellite communication systems that rely on complex ground-station infrastructure and user terminals connected via gateways, DTC communications have a simpler system architecture, with direct links between user devices and satellites, thereby improving overall operational efficiency. 
In addition, using the DTC communication model can overcome the limitations of traditional satellite systems, such as high latency and limited coverage \cite{zhang2023intelligent}.
This enables seamless connectivity for various scenarios, including machine-type communications (MTC), hotspot areas, and the Internet of Things (IoT).

For a single satellite, the scarcity of spectrum resources remains a critical issue in DTC communications, particularly as the number of users increases.
Considering this requirement and the need for more efficient spectrum utilization, traditional orthogonal multiple access (OMA) schemes may struggle to meet this growing demand.
This has drawn the attention of both academic and industry researchers to non-orthogonal multiple access (NOMA) schemes, in which superposed power-domain signals are transmitted, and successive interference cancellation (SIC) is applied at the receiver to decode the superposed signals and remove the interference.
The main advantage of NOMA in DTC (NOMA-DTC) communications is its ability to support massive connectivity. 
NOMA enables simultaneous communication among multiple devices on the same time-frequency resource block (RB), with users differentiated by their power levels. 
Herein, NOMA-based power allocation can be adjusted to meet varying space-ground channel conditions. Compared to OMA schemes, more efficient service provision can be offered. 
Moreover, NOMA can reduce latency in DTC communications. 
Multiple users are served concurrently by a satellite, eliminating the need to wait for dedicated time slots. 
This applies to time-sensitive applications, e.g., real-time data streaming, remote sensing, and emergency communications.

\begin{figure*}[tp]
\begin{center}
\setlength{\abovecaptionskip}{+0.2cm}
\setlength{\belowcaptionskip}{-0.0cm}
\centering
  \includegraphics[width=7.0in, height=3.5in]{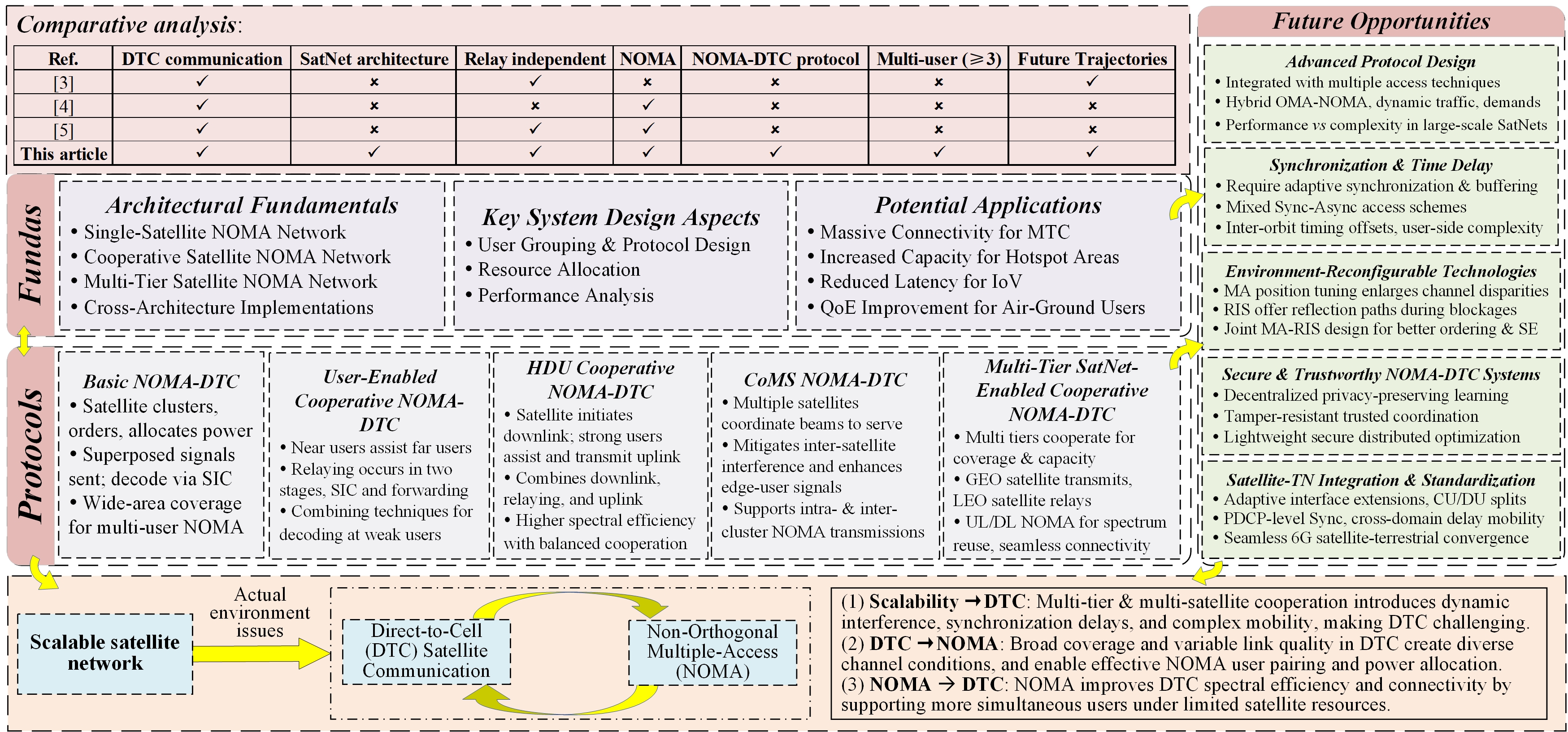}
\renewcommand\figurename{FIGURE}
\caption{\scriptsize Comparative analysis and overview of NOMA for satellite DTC communication systems.}
\label{fig:table}
\end{center}
\vspace{-8mm}
\end{figure*}

Initially, devices such as smartphones could connect to satellites via relayed device-to-device (D2D) communications. In recent years, many devices have been equipped with direct-to-satellite functionality to expand service options. 
A survey \cite{pasandi2024survey} focused on advances, challenges, and prospects of DTC communications but did not consider the benefits of integrating NOMA.
The authors of \cite{yan2018outage} demonstrated that a satellite can serve TN users via NOMA to achieve better outage performance, but a relay is required to support the network.
In \cite{dong2025outage}, a two-user NOMA-DTC scenario was explored to demonstrate lower error rates.
The above works did not comprehensively examine the protocols and future research pathways of NOMA-DTC communications. Moreover, they lack a thorough treatment of architectural foundations, system design, and potential applications of emerging satellite networks (SatNets).
On this basis, a comparison of relevant works and an overview of NOMA-DTC communications are presented in this article in Fig. \ref{fig:table}.

\captionsetup{font={scriptsize}}

Unlike existing articles that either focus on DTC communications without multiple-access enhancement or study NOMA in isolated satellite scenarios, this article provides the first protocol-oriented and architecture-aware overview of NOMA-enabled DTC communications under evolving SatNet paradigms.
Specifically, we study architectural fundamentals, protocol design, and performance insights in a unified framework for single-satellite, cooperative, and multi-tier SatNets.
The main contributions are summarized as follows:
\begin{itemize}
    \item \textbf{Architecture-aware NOMA-DTC framework:} We systematically classify NOMA-DTC under three representative SatNet architectures, i.e., single-satellite NOMA, cooperative satellite NOMA, and multi-tier satellite NOMA networks, and examine their fundamentals and architectural designs. Key system design aspects and potential applications are then explored for practical scenarios.
    \item \textbf{Protocol-level taxonomy and operational workflow:} We investigate five potential protocols of NOMA-DTC under different SatNet architectures and user cooperative modes, e.g., basic single-satellite NOMA-DTC, user-enabled cooperative NOMA-DTC, coordinated multi-satellite NOMA-DTC, hybrid downlink and uplink cooperative NOMA-DTC, and multi-tier SatNet-enabled cooperative NOMA-DTC. 
    \item \textbf{Typical case analysis and future visions:} To show the benefits of NOMA schemes, we study the basic and cooperative NOMA-DTC protocols in a typical case study to show their enhanced spectral efficiency (SE) over OMA schemes. Opportunities for potential research are also outlined for NOMA-DTC communications.
\end{itemize}

\section{Fundamentals of NOMA-DTC}
In this section, we first introduce three architectural fundamentals for NOMA-DTC communications considering evolving SatNet architectures.
Then, we discuss system design aspects from three perspectives.
In addition, we explore several potential applications and their use cases for real-world implementations.

\subsection{Architectural Fundamentals}
\begin{figure}[tp]
\begin{center}
\setlength{\abovecaptionskip}{+0.2cm}
\setlength{\belowcaptionskip}{-0.0cm}
\centering
  \includegraphics[width=3.4in, height=3.2in]{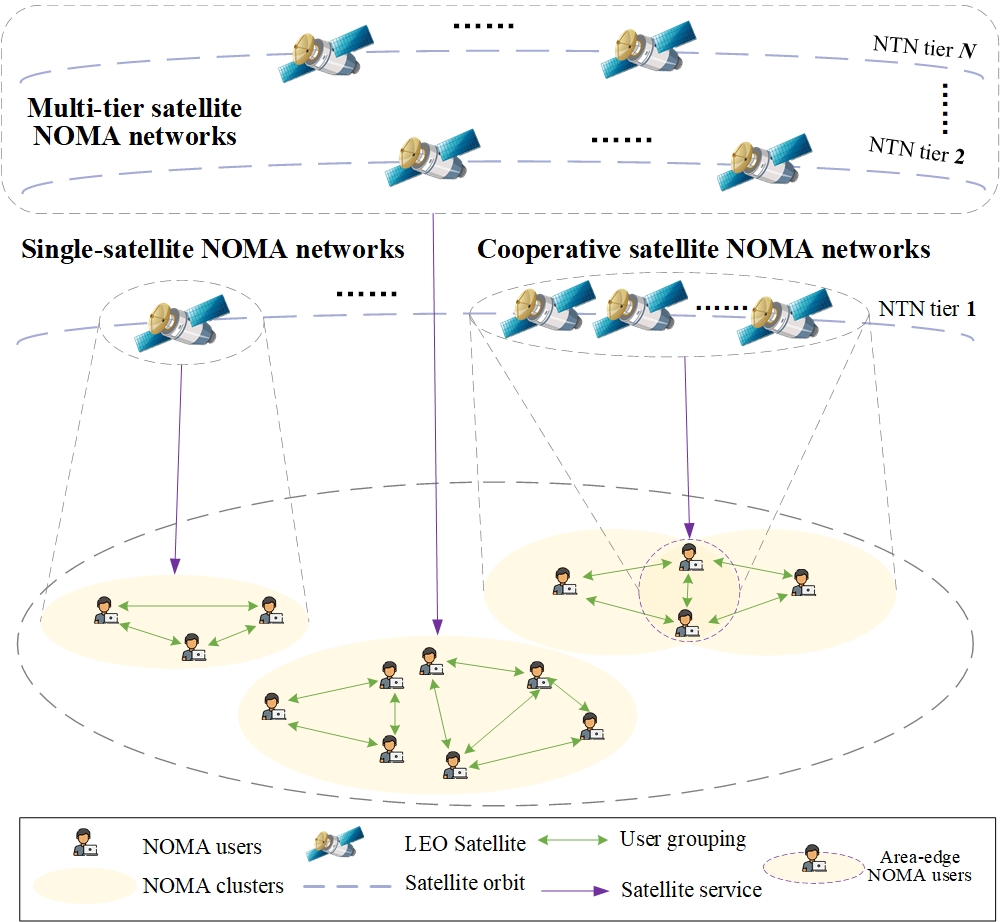}
\renewcommand\figurename{FIGURE}
\caption{\scriptsize Architectures of SatNets for NOMA-DTC communications.}
\label{fig_Arch_Designs}
\end{center}
\vspace{-8mm}
\end{figure}

Although NOMA has been investigated in cases where a single satellite serves two NOMA users to improve their overall performance, its suitability for future SatNets remains unexplored.
Here, we first review the application of NOMA in a single-satellite-supported DTC scenario. Then, considering two potential future SatNets, we discuss how NOMA fits into each network setup, as shown in Fig. \ref{fig_Arch_Designs}.

\captionsetup{font={scriptsize}}

\subsubsection{Single-Satellite (SglS) NOMA Network}
Currently, each satellite has its own ground service area, where users are randomly distributed. TNs cannot serve these users due to their remote locations and the inefficiency of TN infrastructure. 
Using the SIC technique, each satellite identifies the number of users in its serving area and then transmits the data symbols to each user in the same time-frequency RB.
Note that, due to the current association policy, these users have access to this single satellite for service provision only, and no other satellite-serving link connections are available.

Most previous work, e.g., \cite{dong2025outage}, assumes a two-user scenario for simplicity in terms of performance analysis, resource allocation, and practical aspects such as imperfect channel state information (CSI). 
However, when a large number of users need to be served, it is imperative to assess the feasibility of relevant algorithms and technical policies in multi-user scenarios with evolving SatNets.

Moreover, while single-satellite NOMA provides a fundamental framework, its performance may be limited by coverage constraints and inter-satellite interference in dense constellations. This motivates the introduction of cooperative satellite architectures.

\subsubsection{Cooperative Satellite NOMA Network}
By employing the coordinated multi-satellite (CoMS) joint transmission strategy and providing multiple desired signal links to NOMA users from different satellites, the NOMA-DTC system is expected to achieve expanded coverage and improved data rates.
Then, the NOMA users served will also experience relatively less inter-satellite interference.

Define \textit{CoMS-NOMA set} as the set of satellites that cooperate to serve NOMA users. The term \textit{CoMS-NOMA user} refers to a NOMA user that is jointly served by satellites in a \textit{CoMS-NOMA user}, 
while a \textit{non-CoMS-NOMA user} refers to a NOMA user in the same area as a \textit{CoMS-NOMA user} but is only served by one satellite in the \textit{CoMS-NOMA set}.
When multiple NOMA users are served by a satellite, the quality of experience (QoE) of some interference-prone users on the edge of the area needs to be enhanced. With aggregated desired signals from multiple satellites, the performance of these \textit{CoMS-NOMA users} can be much higher than that of non-\textit{CoMS-NOMA users}.

Beyond horizontal cooperation among satellites at the same altitude, vertical cooperation across different orbital layers further enhances flexibility and coverage, leading to the concept of multi-tier satellite NOMA networks.

\subsubsection{Multi-Tier Satellite NOMA Network}
Due to the launch of satellites to different altitudes, SatNets formed by selecting satellites from different altitudes are referred to as ``multi-tier SatNets'' \cite{okati2023stochastic}.
Given the availability of satellites in these tiers, selecting the most appropriate satellite for single-satellite connectivity requires careful exploration.
Furthermore, to facilitate cooperation among satellites from different tiers, multi-tier satellite clustering strategies should be designed to demonstrate how to select satellites at different altitudes.

In practical multi-tier satellite NOMA deployments, environmental factors such as orbit perturbation and atmospheric attenuation can be addressed or mitigated by predictive channel modeling, adaptive power/rate control, and spatial diversity across orbital layers. Strategies such as real-time feedback and learning-based compensation can further enhance the robustness of NOMA-DTC communications under dynamic space conditions.

\subsubsection{Cross-Architecture Implementations}
The architectural designs described above inherently rely on coordination and information exchange among satellites. In practice, such coordination is supported by inter-satellite links (ISLs), whose latency and bandwidth limitations may affect real-time decision-making, particularly for dynamic user grouping and NOMA power allocation.

Excessive signaling overhead or delayed CSI exchange may reduce coordination capabilities. Therefore, practical implementations may adopt semi-static resource allocation, cluster-level coordination, or predictive channel estimation to mitigate ISL dependency. Optical inter-satellite links (OISLs) have also received much research attention to provide ultra-high bandwidth, ultra-low latency links for inter-satellite communications and coordination.

\captionsetup{font={scriptsize}}
\begin{figure*}[tp]
\begin{center}
\setlength{\abovecaptionskip}{+0.2cm}
\setlength{\belowcaptionskip}{-0.0cm}
\centering
  \includegraphics[width=6.4in, height=4.0in]{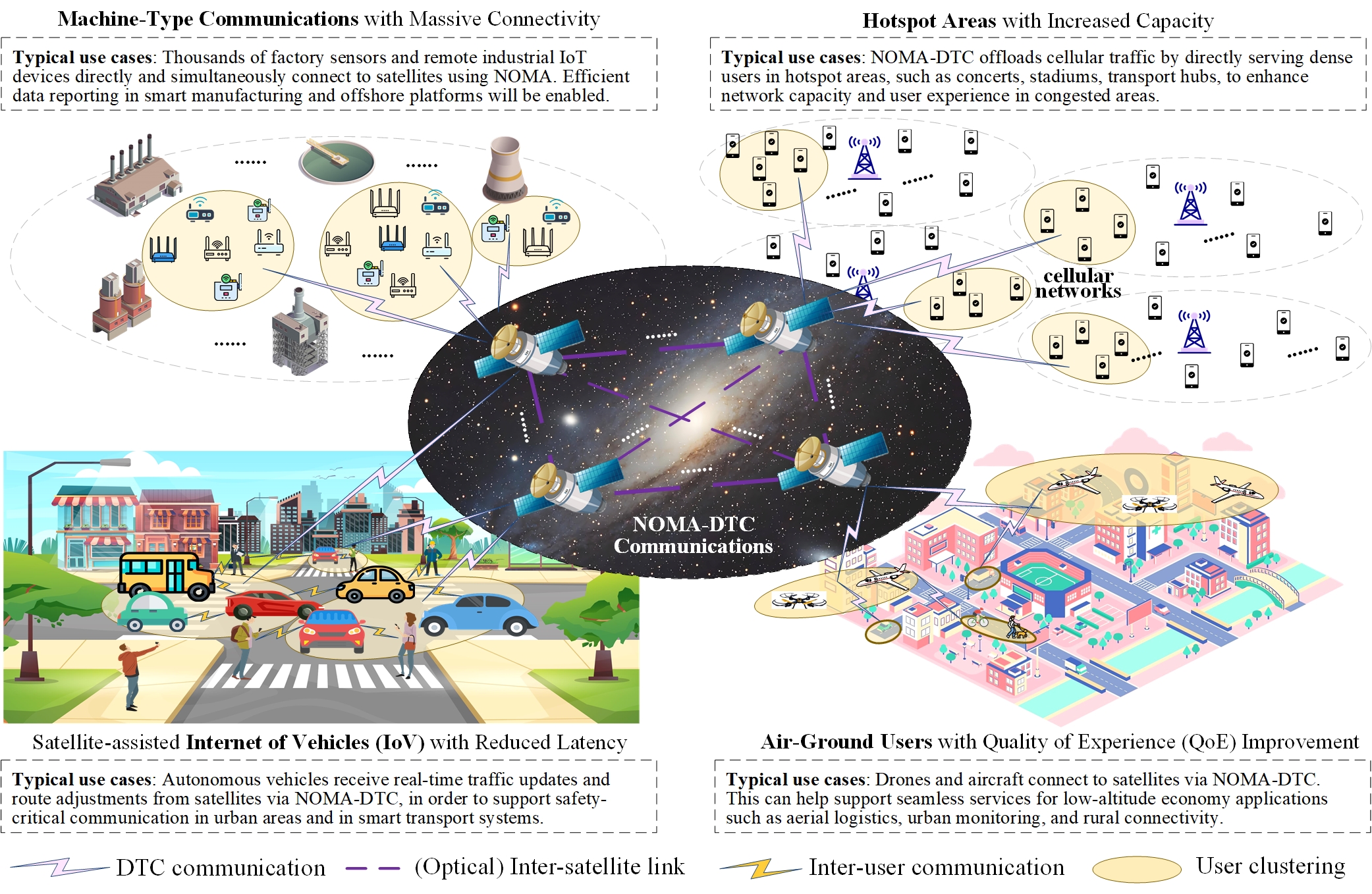}
\renewcommand\figurename{FIGURE}
\caption{\scriptsize Conceptual vision of potential applications in NOMA-DTC communications, including: (1) massive machine-type communications with massive connectivity, (2) increased network capacity for hotspot areas, (3) satellite-assisted IoV with reduced latency, (4) improvement of QoE for air-ground users and for the low-altitude economy.}
\label{fig_Applications}
\end{center}
\vspace{-8mm}
\end{figure*}

\subsection{Key System Design Aspects}
Since TN infrastructure, such as base stations and relay nodes, is not involved, the performance improvement of NOMA-DTC systems largely depends on protocol designs, resource allocation schemes, and proper verification on satellites. However, they may face the following key aspects.

\subsubsection{User Grouping and Protocol Design}
Effective user grouping and protocol design are crucial for achieving a balance between SE and user fairness. 
Typically, to enable power-domain multiplexing and SIC, users with significantly different channel conditions are grouped. Generally, grouping strategies should account for differences in satellite-user channel capacity, user locations, and service demands. 
To balance SE and user fairness, representative metrics such as proportional fairness and max–min rate optimization can be considered. 
For instance, proportional fairness provides a compromise between throughput maximization and fairness, while max–min rate ensures service guarantees for edge users. These metrics can guide adaptive power allocation and user grouping in practical NOMA-DTC systems.
On the other hand, protocols must evolve beyond the commonly used two-user assumptions, which are primarily designed for dynamic group sizes and flexible decoding orders. Moreover, user cooperation, such as relaying and edge-aware beam management, can further improve system SE, especially when users are located at coverage boundaries. These aspects require robust and adaptive protocol mechanisms for practical implementations of NOMA-DTC communications.

\subsubsection{Resource Allocation}
Effective resource allocation in NOMA-DTC communications is crucial for maximizing system SE while ensuring fairness and reliability for multiple users. 
Unlike the traditional OMA scheme, the NOMA scheme requires careful management of time, frequency, and power resources to ensure that users with different channel conditions can share the same resources without excessive interference.
Specifically, regarding power allocation among users, although it is intuitive to give higher service priority and allocate more power to stronger users with better channel conditions to improve the overall system SE, a trade-off between network-wide metrics and individual user experiences should be struck to guarantee the performance of weaker users.

From an optimization perspective, the SE–fairness trade-off can be interpreted as a power allocation problem within each NOMA cluster. The objective is to maximize the aggregate SE of all multiplexed users, while satisfying three key constraints: (i) a total satellite transmit power budget, (ii) minimum rate requirements to guarantee fairness for weaker users, and (iii) decoding feasibility conditions to ensure successful SIC operation.
Allocating more power to strong users typically improves the network-wide sum SE but may degrade the achievable rates of weaker users. Conversely, enforcing fairness-oriented allocation reduces the achievable sum SE. Therefore, practical NOMA-DTC systems require adaptive power control mechanisms that balance overall throughput and user-level service qualities.

\subsubsection{Performance Analysis}
Two key components are discussed to fully reflect the achievable performance of the NOMA-DTC system: modeling TN users and NTN satellites, and characterizing inter- and intra-satellite interference.
Firstly, given the mobility of satellites and the distribution of ground users, appropriate models that reflect the real-time positions of satellites, users, and their channels lay a solid foundation for approximating the proposed mathematical models in practical implementations.
Secondly, owing to the development of satellite mega-constellations in recent years, a satellite may serve multiple users in the same time-frequency RB, so not only inter-satellite but also intra-satellite interference should be taken into account. 

The impact of intra-satellite interference can be viewed as a reduction in the satellite's transmit power allocated to a user. 
Then, the effective transmit power is the portion of the transmit power remaining after accounting for intra-satellite interference.
To ensure NOMA operation, the effective transmit power should be greater than 0, which is referred to as \textit{NOMA necessary condition} \cite{ali2019downlink,li2026coverage}. 
Specifically, this condition reflects a fundamental decoding requirement: the power allocated to each user must be larger than the intra-satellite interference generated by signals intended for other users within the same RB.

As the SINR threshold increases or as more users are multiplexed, the cumulative intra-satellite interference may exceed the allocated power portion for certain users. Once this condition is violated for any user, the effective signal required for successful SIC becomes insufficient, leading to decoding failure for this user and, consequently, the breakdown of the entire NOMA cluster. Then, NOMA operation becomes infeasible for the entire group and all users are assumed to be out of coverage. This explains the sharp performance “collapse” observed in the subsequent case study.

\captionsetup{font={scriptsize}}
\begin{figure*}[tp]
\begin{center}
\setlength{\abovecaptionskip}{+0.2cm}
\setlength{\belowcaptionskip}{-0.0cm}
\centering
  \includegraphics[width=7.0in, height=3.8in]{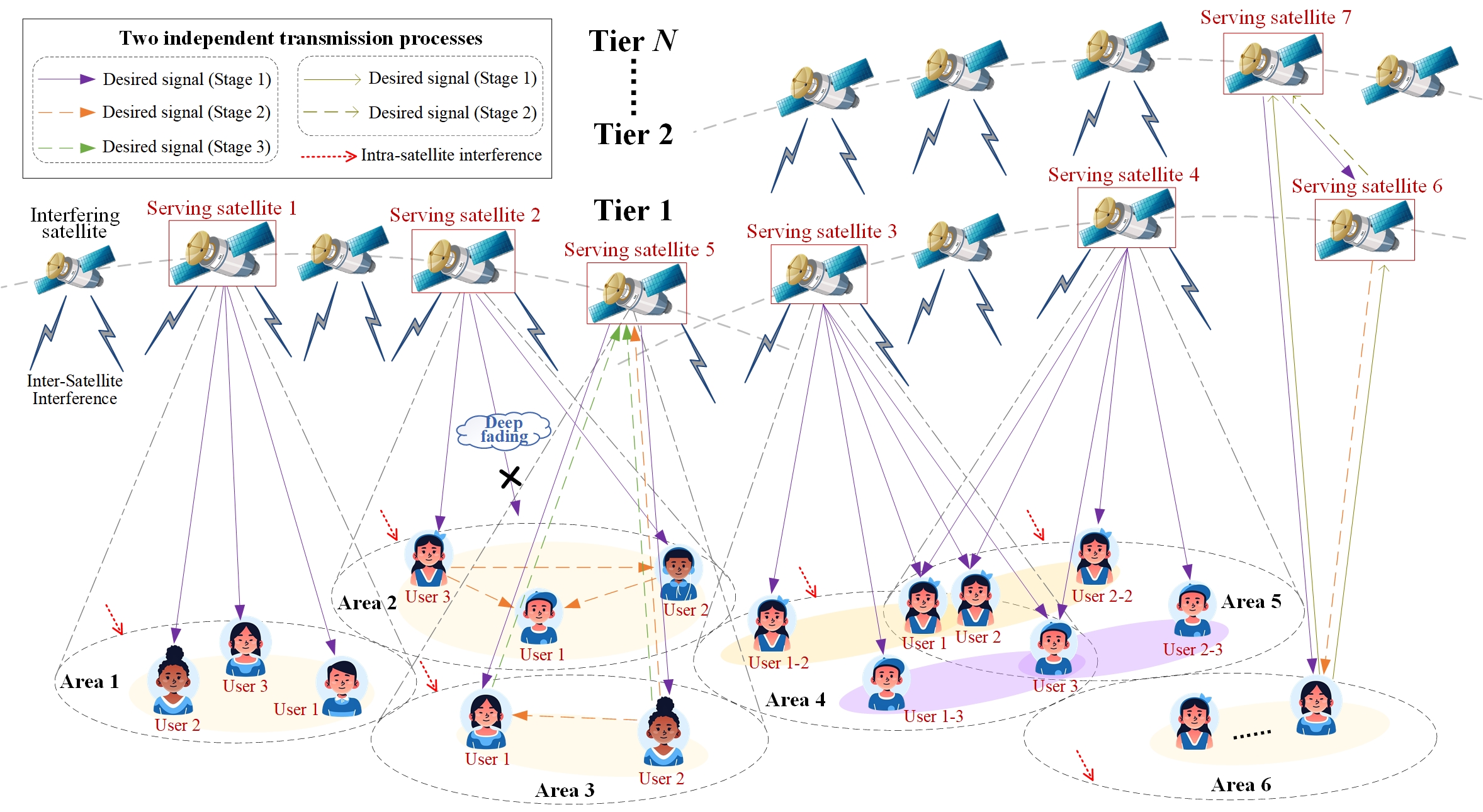}
\renewcommand\figurename{FIGURE}
\caption{\scriptsize Five protocols of NOMA-DTC communications in large-scale global satellite networks: \textbf{Area 1} shows the operation of \textit{basic NOMA-DTC}, \textbf{Area 2} shows the operation of \textit{user-enabled Cooperative NOMA-DTC}, \textbf{Area 3} shows the operation of \textit{HDU cooperative NOMA-DTC}, \textbf{Area 4 and 5} shows the operation of \textit{CoMS-NOMA-DTC}, \textbf{Area 6} shows the operation of \textit{multi-tier SatNet-enabled cooperative NOMA-DTC}. As shown in the legend in the upper left corner, solid and dotted lines are marked in colors to indicate different types of signals and different transmission stages.}
\label{fig_Protocols}
\end{center}
\vspace{-8mm}
\end{figure*}

\subsection{Potential Applications}
In this subsection, we introduce four applications for NOMA-DTC communications, as depicted in Fig. \ref{fig_Applications}.

\subsubsection{Massive Connectivity for MTC}
In recent years, a growing number of devices have been observed in outdoor MTC scenarios such as smart manufacturing and offshore platforms. Traditional OMA schemes are limited by their restricted coverage and access capacity. However, by enabling broader coverage through DTC links and applying NOMA schemes, satellite power levels can be allocated to each user to accommodate more users. This can enhance spectrum utilization and provide scalable and reliable connectivity for devices across various industries.

\subsubsection{Increased Capacity for Hotspot Areas}
In hotspot areas where user density is high, such as stadiums or transport hubs, DTC communications often face congestion and capacity limits. By using NOMA in such situations, simultaneous access can be supported for high-demand users, enabling more efficient use of available spectrum.
This integration will significantly improve the overall capacity of the network in urban environments and large-scale public events.

\subsubsection{Reduced Latency for Internet-of-Vehicles}
NOMA-DTC communications can reduce latency in satellite-assisted Internet-of-Vehicles (IoV) applications, such as autonomous driving on city roads and highways, where low latency is required to ensure real-time communication between vehicles and satellites. 
By allowing multiple vehicles to transmit and receive data simultaneously on the same time-frequency RB, NOMA reduces access delays caused by dedicated time slots. This will support real-time traffic management, autonomous driving, and vehicle-to-everything (V2X) communications.

\subsubsection{QoE Improvement for Air-Ground Users}
Air-ground users with satellite connections often experience variable space-ground channel conditions due to high mobility and changing atmospheric conditions. For airborne users, such as aircraft and unmanned aerial vehicles (UAVs) in low-altitude economy scenarios, NOMA allows simultaneous data transmission and user prioritization based on their channel quality. 
This improves spectrum efficiency and ensures more reliable connectivity for services such as video streaming, navigation, and emergency communications.

\section{Protocols of NOMA-DTC}
This section focuses on the protocols for NOMA-DTC communications to enhance system analysis and improve user performance through user collaboration and satellite coordination, as shown in Fig. \ref{fig_Protocols}.

\subsection{Basic NOMA-DTC}
Intuitively, basic NOMA-DTC aims to enhance SE by allowing multiple users within a satellite beam to share the same time-frequency RB through power-domain multiplexing. The core idea is to exploit channel disparities among users so that SIC can be efficiently applied.

\begin{figure*}[tp]
\begin{center}
\setlength{\abovecaptionskip}{+0.2cm}
\setlength{\belowcaptionskip}{+0.2cm}
\centering
  \includegraphics[width=7.0in, height=2.3in]{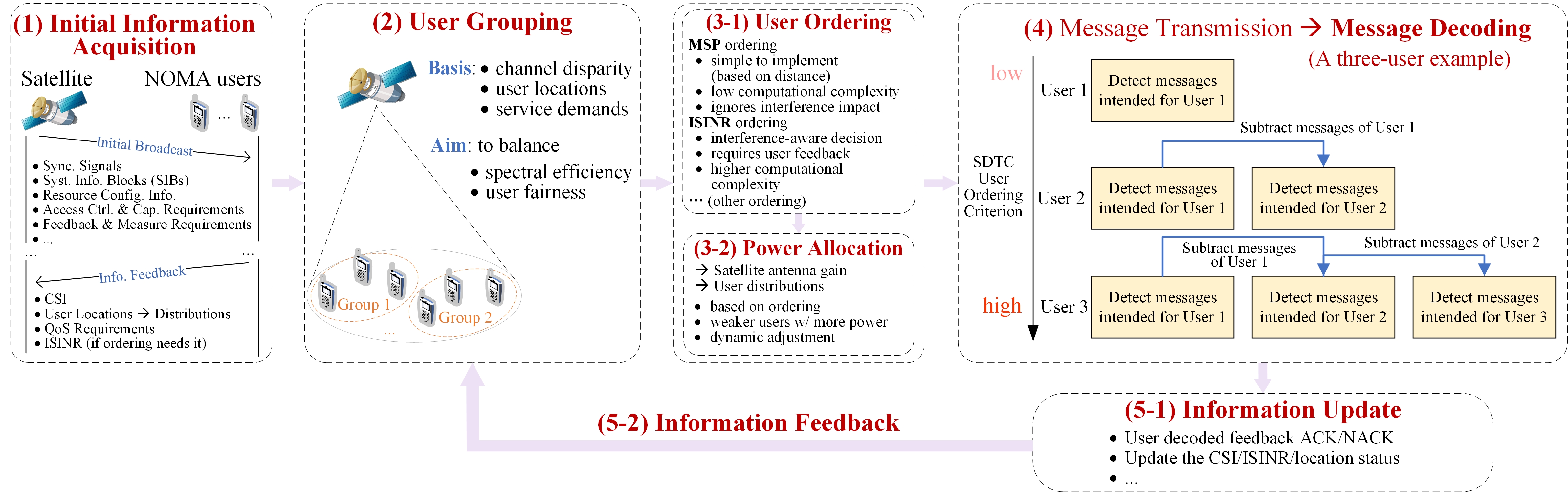}
\renewcommand\figurename{FIGURE}
\caption{\scriptsize An illustration of the end-to-end NOMA-DTC communication process. (1) Initially, the satellite broadcasts control and configuration information to users. Then, NOMA users feed back CSI, QoS, ISINR (if required), etc. (2) Based on user locations and channel disparities, users are grouped into different clusters, each served by a beam of the satellite. (3) Focusing on a certain beam, a selected scheme orders users, and the power allocation is adapted accordingly. (4) During message transmission, SIC is applied at the receivers to decode intended signals, and a three-user decoding process is provided. (5) Finally, related information, including CSI, ISINR, and decoding status, is updated and returned as feedback to the system. These steps form a closed-loop design for robust and efficient NOMA-DTC communications.}
\label{fig_DTC_NOMA_process}
\end{center}
\vspace{-8mm}
\end{figure*}

In a basic NOMA-DTC scenario, as shown in \textbf{Area 1} of Fig. \ref{fig_Protocols}, a satellite uses NOMA to allow multiple users in its serving area to share the same time-frequency RB.
Specifically, an overview of the exemplar end-to-end NOMA-DTC communication process is presented in Fig. \ref{fig_DTC_NOMA_process}.
After obtaining initial information from NOMA users, the serving satellite groups users into clusters, each potentially containing more than two users. In each group, users are ordered according to the designated criterion and then assigned transmit power. The related information will be updated periodically. The above process is fundamental and can be extended to cooperative, multi-tier satellite-based NOMA-DTC communications.

Before transmissions, the satellite first decides the users' ordering. 
Multiple signals are superposed on the satellite and transmitted to users, who then apply SIC to decode the superposed signals and remove the interference caused by superposition. Specifically, a three-user example of message decoding is given in Step (4) of Fig. \ref{fig_DTC_NOMA_process}.
A user with a larger user index, i.e., User 3, is the strongest user, while a user with a smaller user index, i.e., User 1, is the weakest user. 
User 1 treats messages for Users 2 and 3 as noise and detects and decodes its own messages. 
User 2 uses SIC to detect, decode, and subtract messages for User 1, before detecting and decoding messages for itself. 
User 3 also uses SIC to sequentially detect, decode, and subtract messages for User 1 and User 2, before detecting and decoding messages for itself. 
Following these principles, message decoding for more than three users can be designed subsequently.

In general, users' ordering may depend on the channel quality or instantaneous signal-to-interference-plus-noise ratio (ISINR); for simplicity, it may sometimes be decided by the satellite-user distances, i.e., mean square power (MSP) to users. 
This is because small-scale fading may not affect channel state information (CSI) much over long distances \cite{ali2019downlink}. 
On this basis, MSP ordering is denoted as a fading-free ordering that uses the total unit power received at users, which can then be equivalent to the order of satellite-user distances; ISINR ordering is denoted as a fading-based ordering that involves ISINR at users \cite{li2026coverage}.
Considering a satellite's larger serving area, NOMA users would be equipped with higher decoding and detecting capabilities, allowing more NOMA users to be served simultaneously.

While extending NOMA-DTC beyond the traditional two-user assumption enhances SE, it inevitably increases the computational burden associated with SIC. Specifically, each additional multiplexed user introduces extra decoding and interference subtraction stages, leading to increased processing latency, memory usage, and energy consumption at the user side. 
Since users typically have more constrained processing capabilities than satellites in practical deployments, high-order NOMA clusters with more than three users may impose non-negligible hardware complexity and power consumption overhead. 
To ensure implementation feasibility, the cluster size in practical DTC systems is expected to remain small, e.g., two or three users per RB, with adaptive user grouping applied based on user capability and service requirements.
To deal with this issue, emerging hardware acceleration techniques, such as parallel decoding architectures and dedicated signal processing units, can partially mitigate SIC-related computational overhead. Reduced-complexity SIC techniques, early termination strategies, and partial interference cancellation schemes may also be adopted to balance performance and complexity.

\captionsetup{font={scriptsize}}

\subsection{User-Enabled Cooperative NOMA-DTC}
The fundamental motivation behind user-enabled cooperative NOMA-DTC is to leverage the stronger channel conditions of near users to assist far users. By introducing relaying among users, both reliability and coverage for edge users can be improved.

In the cooperative NOMA-DTC protocol, when the serving satellite sends data to multiple users, they help each other improve mutual performance and overall network performance. 
This cooperation mode mainly takes the form of user relaying, where the source satellite sends data to all users, where near users can act as relays for far users.
There are two stages to complete the data transmission, with an example of cooperative NOMA between three users, as shown in \textbf{Area 2} of Fig. \ref{fig_Protocols}. The operations are shown as follows.

In the first stage, the source satellite sends the superimposed signals to three users. The nearest satellite user, i.e., User 3, acts as a relay for User 2 and User 1. It first decodes the signals intended for the other two users, considering its own signal as noise. Then, User 3 uses SIC to cancel the signals intended for User 2 and User 1, and obtains its own signal. Similarly, User 2 acts as a relay for User 1. 
When a direct link exists between the satellite and the farthest User 1, User 1 treats the signals intended for User 3 and User 2 as noise and decodes the signal for itself.
When there is no direct link between the satellite and User 1, User 1 does not receive effective signals at this stage.
Note that the existence of a “direct-link” between the satellite and the far user depends on practical channel conditions and blockage events. In operational systems, the direct link can be considered available when the received signal quality satisfies a predefined decoding threshold\footnote{Otherwise, the system adaptively switches to the relay-assisted second-stage transmission. This threshold-based switching mechanism enables practical deployment of the two-stage cooperative NOMA-DTC protocol under dynamic space–ground channel conditions.}. Otherwise, the link is regarded as unreliable or blocked.

In the second stage, both User 3 and User 2 forward the signal intended for User 1. If signals from both satellites and other users are received at both stages, User 1 will use different techniques, such as maximal-ratio combining, equal-gain combining, and selected combining, to detect signals \cite{zeng2020cooperative}.

\subsection{Hybrid Downlink-Uplink (HDU) Cooperative NOMA-DTC}
By integrating both downlink and uplink processes for satellite users, an HDU cooperative NOMA-DTC scheme can be realized. 
This scheme leverages both downlink and uplink communication channels in a cooperative and concurrent manner \cite{wei2017performance}. 
Compared with cooperative NOMA-DTC, strong users not only relay messages from their local terminals to assist weak users, but also conduct uplink transmissions to the serving satellite. As an example of this protocol with two users, there are three stages, as shown in \textbf{Area 3} of Fig. \ref{fig_Protocols}. 

The first stage is for the NOMA-DTC downlink. The satellite transmits the superimposed signal, which consists of two signals that respectively target two users, i.e., User 1 and User 2.
The second stage is for cooperative transmission. The stronger User 2 broadcasts the superimposed signal, which consists of one signal obtained from the SIC process for User 2, and another signal from its own for uplink transmission to the satellite.
The third stage is for NOMA-DTC uplink. Both User 1 and User 2 transmit their uplink signals to the satellite simultaneously.
The HDU cooperative NOMA-DTC scheme is expected to promote efficient resource sharing and mitigate interference, thereby enhancing the sum SE of the system, although at the cost of a slight reduction in signal reception reliability for weaker users.

\subsection{CoMS NOMA-DTC}
The benefit of CoMS NOMA-DTC is to transform inter-satellite interference into useful signal components through coordinated transmission. Instead of serving users independently, multiple satellites jointly enhance the received signal strength for selected users.

To maximize the sum rate for downlink transmission with NOMA-DTC, satellite power allocation enables NOMA users to perform SIC according to the ascending order of their channel gains. 
Before decoding the desired signal, each user will cancel the signals of other users with channel gains lower than those of the considered user \cite{ali2018coordinated}. As a result, users at the edge of the serving area generally experience intra-satellite interference due to signals for NOMA users at the center.
On the other hand, to ensure an optimal power allocation strategy, a lower channel gain user will have a lower SINR because they are more prone to inter-satellite interference (ISI).
To mitigate such ISI for NOMA users, it is crucial to consider the CoMS NOMA-DTC protocol, where multiple satellites and their associated beams are jointly coordinated to serve some of the users, which are referred to as CoMS users, i.e., User 1, User 2, User 3  in \textbf{Areas 4 and 5} of Fig. \ref{fig_Protocols}. 

Note that CoMS-NOMA users can be within the same NOMA cluster or across different clusters. When two COMP-NOMA users, User 1 and User 2, are in the same NOMA cluster (in orange) with non-CoMS-NOMA users (i.e., User 1-2, User 2-2), both receive signals from the two serving satellites simultaneously over the same time-frequency RB. 

Another User 3 is also served by these two satellites, but is in a separate NOMA cluster (in purple) with non-CoMS-NOMA users (i.e., User 1-3, User 2-3), so that the signals it receives are orthogonal to those of User 1 and User 2.
The integration of CoMS with NOMA-DTC communications can improve the quality of the received signal while reducing inter-beam or inter-satellite interference \cite{li2024analytical}.

\subsection{Multi-Tier SatNet-Enabled Cooperative NOMA-DTC}
Multi-tier SatNet-enabled cooperative NOMA-DTC operates across multiple constellations or satellite layers, such as LEO, MEO, and GEO, to improve coverage, capacity, and service continuity.
In this protocol, a satellite may act as a transmitter or a relay, thus developing a form of cooperative relaying for NOMA-DTC, as shown in \textbf{Area 6} of Fig. \ref{fig_Protocols}.

Consider a GEO satellite as a transmitter and an LEO satellite as a relay. For downlink transmissions, in the first stage, the GEO satellite first broadcasts superimposed signals not only to users but also to another LEO satellite located at a lower altitude. The signal strength received and decoded by a user may not be high enough. 
Then, in the second stage, the LEO satellite forwards the decoded signals to the destination users. 
In addition, uplink transmissions from a user to both an LEO satellite and a GEO satellite using NOMA can be considered, especially in times of inline interference, i.e., the LEO satellite falls into the line-of-sight (LoS) path between the GEO satellite and the user \cite{ge2021performance}.
This multi-tier structure enables efficient spectrum reuse and seamless handovers between satellite layers, providing uninterrupted connectivity.

\captionsetup{font={scriptsize}}
\begin{figure*}[tp]
\begin{center}
\setlength{\abovecaptionskip}{+0.2cm}
\setlength{\belowcaptionskip}{-0.0cm}
\centering
  \includegraphics[width=6.8in, height=3.2in]{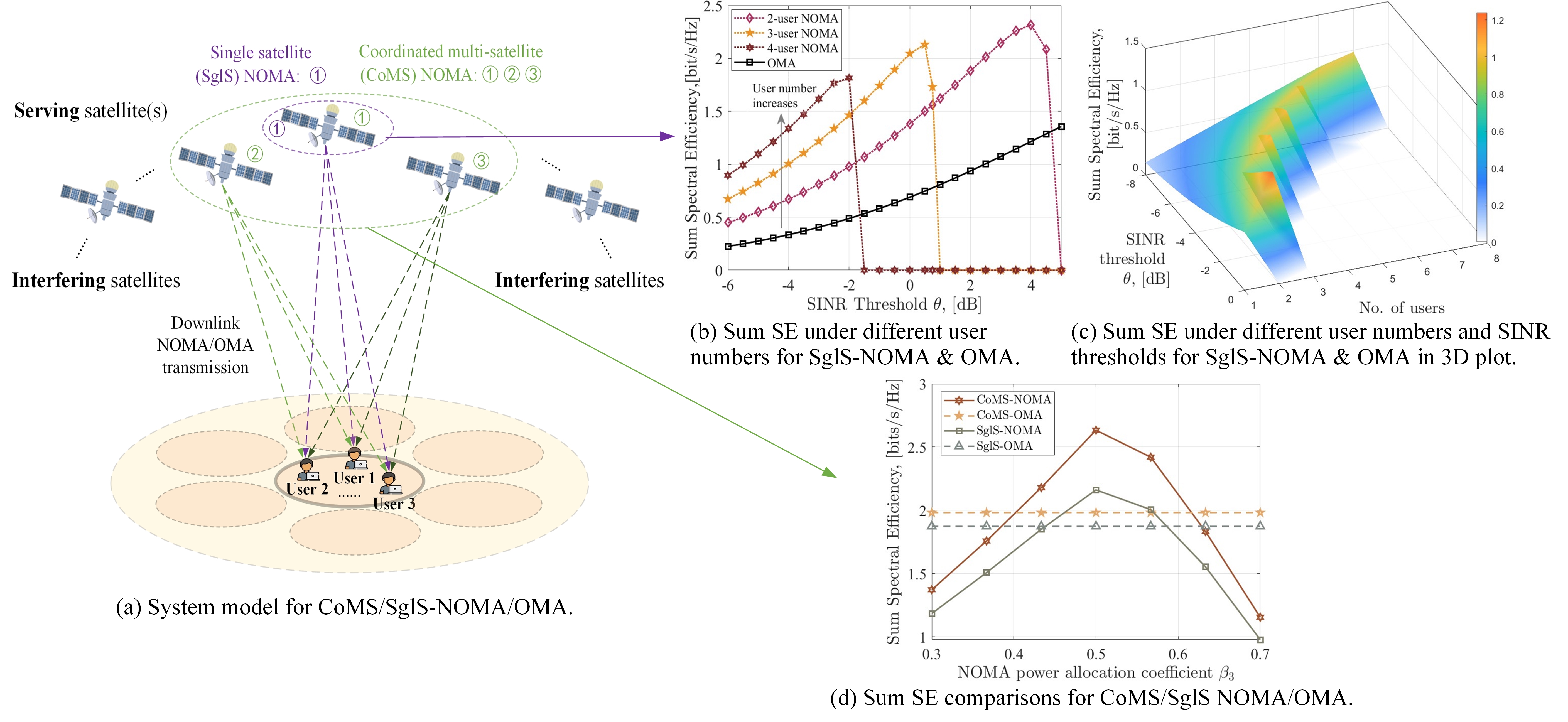}
\renewcommand\figurename{FIGURE}
\caption{\scriptsize 
A typical system model where a single satellite or multiple cooperative satellites are designated to simultaneously serve multiple users in a target area. In this setup, both large-scale fading with a carrier frequency of $f_c=2$ GHz and Nakagami small-scale fading are considered, where the fading parameter $m=2$ is selected to reflect LoS components. 
The radius of Earth is $R_\text{E}=6,371.393$ km, and a total number of $600$ LEO satellites are distributed at the altitude $H_\text{N}=500$ km. The transmit power, main-lobe gain, and side-lobe gain are $P_\text{N}=50$ dBm, $G_\text{ml}=30$ dBi, and $G_\text{sl}=10$ dBi, respectively. NOMA users are distributed in a circular serving area of a satellite with a radius $R_\text{T}=200$ km. A slightly imperfect SIC effect is considered for practical implementation, and the residual SIC factor is $0.1$. The noise power on the user side is $\sigma^2 = -173$ dBm/Hz, and the space-ground path loss exponent is $\alpha = 2.0$.}
\label{fig_Experiment}
\end{center}
\vspace{-6mm}
\end{figure*}

\section{Case Study: Exploiting NOMA for Enhancing DTC Communications}

This case study explores the advantages of SglS NOMA and CoMS NOMA in augmenting DTC communications, as shown in Fig. \ref{fig_Experiment}(a). 
NOMA users are ordered according to the MSP ordering scheme. The NOMA necessary condition is applied to ensure that all users are served in a NOMA manner. Using NOMA, one satellite in the purple dotted circle independently serve the users; while in another case, three satellites in the green dotted circle cooperatively provide services.
In this case study, we mainly focus on sum SE as a primary performance metric to provide a clear comparison between NOMA and OMA schemes. While other metrics such as fairness, latency, and QoE are also critical in NOMA-DTC system designs, SE serves as a representative indicator of system-level performance and spectrum utilization.

In Fig. \ref{fig_Experiment}(b), we compare the sum SE of NOMA with two, three, and four users against that of the SglS NOMA benchmark. Clear performance gains over OMA are observed, and the sum SE increases as more users are multiplexed. However, for each user configuration, the sum SE eventually drops to zero once the SINR threshold exceeds a certain value.
This follows from the previously discussed NOMA necessary conditions: the effective transmit power of each user depends on both its allocated power and the intra-satellite interference it experiences. As the SINR threshold increases, intra-satellite interference can exceed the allocated power to a user, leading to decoding failures and the breakdown of NOMA operation.
Moreover, coverage behavior is also implicitly reflected by such SE collapse, where the violation of decoding conditions leads to zero throughput, effectively corresponding to outage.

For massive connectivity, Fig. \ref{fig_Experiment}(c) compares the sum SE for different numbers of users. 
In practical implementations, fairness can be partially reflected through power allocation constraints and decoding feasibility conditions.
Herein, for fairness and generality, each user is allocated an equal proportion of the satellite’s transmit power. The results show that, for a given SINR threshold, there exists an optimal number of users. 
Specifically, the sum SE increases with the number of users until reaching a peak, beyond which additional users violate the NOMA necessary condition and cause the system performance to collapse to zero.
This reflects an engineering trade-off: whether to admit more users at lower thresholds to enhance connectivity, or to limit the group size at higher thresholds to sustain performance.

Finally, by combining satellite serving modes, i.e., CoMS or SglS, with NOMA or OMA, the sum SE across four schemes is compared in Fig. \ref{fig_Experiment}(d). Three satellites are employed for CoMS operation. Users are ordered by their ascending distance to the area center, with $\beta_3$ denoting the power allocation coefficient for User 3, and $\beta_1=0.2$, $\beta_2=1-\beta_1-\beta_3$ for User 1 and User 2, respectively. CoMS is observed to consistently outperform SglS for both NOMA and OMA. Nevertheless, achieving a performance advantage of NOMA over OMA requires an appropriate choice of NOMA power allocation coefficients\footnote{Notably, the relative performance superiority between CoMS-OMA and SglS-NOMA also depends on the power allocation coefficients. This article presents the performance under one set of power allocation coefficients mainly to provide a comparison between OMA and NOMA.}, highlighting the importance of careful power-domain configuration.

From a practical deployment perspective, results above suggest that NOMA-DTC can significantly enhance SE when user grouping and power allocation are properly configured. However, the existence of an optimal number of multiplexed users indicates that more user admission may degrade performance. Therefore, adaptive clustering strategies that dynamically adjust group size based on SINR thresholds and traffic demand are essential in real-world DTC systems.

Note that the residual SIC factor models imperfect interference cancellation due to practical limitations in channel estimation accuracy, hardware non-ideality, and signal processing precision. This SIC factor typically ranges from 0.01 (worst-case SIC) to 1 (perfect SIC) depending on receiver capability. 
For satellite DTC communications, high Doppler shifts and long propagation delays may further degrade CSI accuracy, which potentially increases residual interference levels. However, with advanced Doppler compensation, predictive channel tracking, and adaptive filtering techniques, moderately imperfect SIC performance remains achievable. In this case study, a residual SIC factor of 0.1 is adopted to reflect a realistic yet non-ideal operating condition.
Although SIC introduces additional processing delay at the receiver, NOMA-DTC reduces access delay by allowing simultaneous transmission without dedicated time-slot scheduling. In satellite systems where propagation delay dominates the latency budget, the incremental SIC processing time is typically negligible compared with the overall round-trip delay, resulting in a net latency benefit for delay-sensitive services.

\section{Future Opportunities}

\subsection{Advanced Protocol Design}
The design of advanced protocols focuses on the interplay between NOMA-DTC and other network-layer technologies. 
For example, network-coded NOMA-DTC can enhance data throughput by exploiting network coding techniques to combine and relay signals. Other multiple-access techniques, such as frequency-division multiple access (FDMA) and code-division multiple access (CDMA), can also be integrated to improve spectrum utilization. In addition, hybrid OMA-NOMA-DTC can be considered for scalable solutions under varying traffic loads. 
A trade-off exists between system complexity and performance improvement. With these advanced calculations and algorithms, computational and communication complexity can increase, and the low-latency requirement may be compromised, particularly in large-scale, dynamic SatNets.
Another important direction is low-complexity SIC design for satellite DTC scenarios, where computational capability, battery constraints, and latency requirements differ significantly from TN base station deployments.

\subsection{Synchronization and Real-Time Implementation}
Synchronization is a critical issue, especially for CoMS NOMA-DTC and multi-tier SatNets. Future research may focus on different synchronization frameworks that tolerate delay misalignment while maintaining decoding accuracy. Adaptive timing control and buffering at the user can help handle out-of-order symbol arrivals. 
Another direction involves the development of synchronous-asynchronous access schemes designed for orbit-specific latency. 
These synchronization mechanisms naturally connect to practical system implementation and enable real-time deployment of NOMA-DTC protocols. Future testbed development, hardware-in-the-loop validation, and emulation of orbit-dependent delays will be essential to verify the feasibility of such synchronization designs in large-scale SatNets. These efforts facilitate the analytical concepts into experimentally validated NOMA-DTC solutions.

\subsection{Emerging Environment-Reconfigurable Technologies}
Future NOMA-DTC systems will rely on reconfigurable environments to achieve reliable performance under dynamic space–ground conditions. 
A movable antenna (MA) offers greater adaptability than fixed-geometry phased arrays, since element positions can be reconfigured to produce flexible beam patterns through joint antenna placement and beamforming optimization. Footprints of LEO satellites shift across regions with varying user densities, but require slowly varying steering. Then, MA position tuning is feasible and can deliberately enlarge channel disparities among users to improve NOMA cluster formation and SIC reliability. 
Complementarily, a reconfigurable intelligent surface (RIS) can introduce controllable reflection paths when blockages occur, and shape received power differences \cite{wu2026intelligent}. 
Therefore, joint MA–RIS design for NOMA-DTC enhances coverage, strengthens channel ordering for NOMA, and boosts SE in DTC scenarios.

\subsection{Secure and Trustworthy NOMA-DTC Systems}
Satellites and users may not fully trust each other due to heterogeneous ownership and limited coordination, especially in multi-satellite cooperative scenarios. To deal with this concern, blockchain-enabled federated reinforcement learning (FRL) offers a promising direction for secure and distributed optimization in NOMA-DTC systems operating under zero-trust environments \cite{mao2025blockchain}. 
Specifically, FRL enables decentralized learning of resource allocation and user association/access policies without sharing raw data, thereby preserving privacy. Meanwhile, blockchain can provide a tamper-resistant ledger to ensure the integrity of model updates, enforce consensus, and prevent malicious behavior. This combination is particularly relevant for dynamic satellite networks, where secure coordination across multiple entities is required. 
Future work may investigate lightweight blockchain protocols and communication-efficient FRL schemes for latency-sensitive satellite NOMA and resource-constrained DTC environments.

\subsection{Satellite-Terrestrial Integration and Standardization}
Future NOMA-DTC systems will require deep integration with terrestrial fifth-generation (5G)/sixth-generation (6G) networks to achieve seamless, scalable hybrid connectivity.
Building upon 3GPP Release 19, possible research topics can include physical-layer design, such as extending interfaces (e.g., Xn-C/Xn-U) between TN and NTN, flexible functional splits between the central unit (CU) and distributed unit (DU), and adaptive signaling procedures for orbital mobility. 
It is also essential to enhance the flexibility of the signal radio bearer (SRB) and data radio bearer (DRB) and enable packet data convergence protocol (PDCP)-level synchronization to support heterogeneous delay and capacity profiles across space-air-ground domains.
Effective interface standardization and protocol will enable efficient satellite-terrestrial convergence in future 6G networks.
In addition, the applicability of NOMA-DTC may vary across frequency bands. In \textit{S}-band systems, wider beam footprints and lower atmospheric attenuation favor massive connectivity scenarios. In contrast, \textit{Ku}- and \textit{Ka}-band systems offer larger bandwidth but experience higher rain attenuation, which may enlarge channel disparities and potentially benefit power-domain multiplexing. Therefore, adaptive NOMA parameter design should consider frequency-dependent propagation characteristics.

\section{Conclusions}
This article investigated the fundamentals, protocols, and future opportunities of NOMA-DTC communications, a crucial technology for next-generation wireless communications.
In particular, we first introduced architectural designs for NOMA-DTC communications that consider different cooperative modes and evolving SatNet architectures, such as cooperative SatNets and multi-tier SatNets. Several key system design aspects and potential applications were also presented.
Then, related protocols for the above cooperative modes and SatNet architectures were presented, followed by a case study demonstrating the effectiveness of NOMA in DTC communications with multiple users compared to OMA. 
Finally, we pointed out the critical and promising future opportunities and hope that this article will be helpful in designing and implementing NOMA-DTC communications.

\section*{Acknowledgment}
The work was supported in part by the Zhejiang Provincial Natural Science Foundation of China under Grant No. LQN25F010003, in part by the Ningbo Natural Science Foundation under Grant 2025J021, in part by the State Key Laboratory of Integrated Services Networks, Xidian University, and in part by the YongRiver Scientific and Technological Innovation Project No. 2023A-187-G.

\bibliographystyle{IEEEtran}
\bibliography{references.bib}

@inproceedings{li2024analytical,
  title={An Analytical Model for Coordinated Multi-Satellite Joint Transmission System},
  author={Li, Xiangyu and Shang, Bodong},
  booktitle={Proc. Int. Conf. Ubiquitous Commun. (Ucom)},
  pages={169-173},
  address={Xi'an, China},
  year={2024},
  month={Jul.},
}

@article{li2026coverage,
  title={Coverage and Spectral Efficiency of {NOMA}-Enabled {LEO} Satellite Networks with Ordering Schemes},
  author={Li, Xiangyu and Shang, Bodong and Wu, Qingqing and others},
  journal={IEEE Trans. Veh. Technol.},
  pages={1--1},
  year={2026},
  month={Jan.},
  publisher={IEEE}
}

@article{okati2023stochastic,
  title={Stochastic Coverage Analysis for Multi-Altitude {LEO} Satellite Networks},
  author={Okati, Niloofar and Riihonen, Taneli},
  journal={IEEE Commun. Lett.},
  year={2023},
  month={Oct.},
  volume={27},
  number={12},
  pages={3305-3309},
  publisher={IEEE}
}

@article{ali2019downlink,
  title={Downlink non-orthogonal multiple access ({NOMA}) in {P}oisson networks},
  author={Ali, Konpal Shaukat and Haenggi, Martin and ElSawy, Hesham and others},
  journal={IEEE Trans. Commun.},
  volume={67},
  number={2},
  pages={1613--1628},
  year={2019},
  month={Feb.},
  publisher={IEEE}
}

@article{zeng2020cooperative,
  title={Cooperative {NOMA}: State of the art, key techniques, and open challenges},
  author={Zeng, Ming and Hao, Wanming and Dobre, Octavia A and others},
  journal={IEEE Netw.},
  volume={34},
  number={5},
  pages={205--211},
  year={2020},
  month={Jul.},
  publisher={IEEE}
}

@inproceedings{wei2017performance,
  title={Performance analysis of a hybrid downlink-uplink cooperative {NOMA} scheme},
  author={Wei, Zhiqiang and Dai, Linglong and Ng, Derrick Wing Kwan and others},
  booktitle={Proc. IEEE 85th Veh. Technol. Conf. (VTC Spring)},
  pages={1--7},
  address={Sydney, NSW, Australia},
  year={2017},
  month={Jun.},
}

@article{ali2018coordinated,
  author={Ali, Md Shipon and Hossain, Ekram and Kim, Dong In},
  journal={IEEE Wireless Commun.}, 
  title={Coordinated Multipoint Transmission in Downlink Multi-Cell {NOMA} Systems: Models and Spectral Efficiency Performance}, 
  year={2018},
  month={Apr.},
  volume={25},
  number={2},
  pages={24-31},
}

@article{bakhsh2024multi,
  title={Multi-satellite {MIMO} systems for direct satellite-to-device communications: A survey},
  author={Bakhsh, Zohre Mashayekh and Omid, Yasaman and Chen, Gaojie and others},
  journal={IEEE Commun. Surveys Tuts.},
  year={2024},
  month={Aug.},
  volume={27},
  number={3},
  pages={1536-1564},
  publisher={IEEE}
}

@article{yan2018outage,
  author={Yan, Xiaojuan and Xiao, Hailin and Wang, Cheng-Xiang and others},
  journal={IEEE Wireless Commun. Lett.}, 
  title={Outage Performance of {NOMA}-Based Hybrid Satellite-Terrestrial Relay Networks}, 
  year={2018},
  month={Jan.},
  volume={7},
  number={4},
  pages={538-541},
}

@article{zhang2023intelligent,
  title={Intelligent channel prediction and power adaptation in {LEO} constellation for 6{G}},
  author={Zhang, Haijun and Song, Wei and Liu, Xiangnan and others},
  journal={IEEE Netw.},
  volume={37},
  number={2},
  pages={110--117},
  year={2023},
  month={Sep.},
  publisher={IEEE}
}

@article{dong2025outage,
  title={Outage performance of {NOMA}-based multi-user satellite communication system under polarization conversion},
  author={Dong, Youran and Xu, Guanjun and Zhao, Nan and others},
  journal={IEEE Trans. Veh. Technol.},
  year={2025},
  month={Mar.},
  volume={74},
  number={3},
  pages={5146-5151},
  publisher={IEEE}
}

@inproceedings{pasandi2024survey,
  title={A survey on direct-to-device satellite communications: Advances, challenges, and prospects},
  author={Pasandi, Hannaneh B and Fraire, Juan A and Ratnasamy, Sylvia and others},
  booktitle={Proc. of Int. Workshop LEO Netw. Commun.},
  pages={7--12},
  address={Washington D.C., DC, USA},
  year={2024},
  month={Nov.}
}

@article{ge2021performance,
  title={Performance analysis of cooperative nonorthogonal multiple access scheme in two-layer {GEO}/{LEO} satellite network},
  author={Ge, Ruixing and Bian, Dongming and An, Kang and others},
  journal={IEEE Syst. J.},
  volume={16},
  number={2},
  pages={2300--2310},
  year={2021},
  month={Jul.},
  publisher={IEEE}
}

@article{wu2026intelligent,
  title={Intelligent reflecting surface for satellite communications: Applications, challenges, and future perspectives},
  author={Wu, Yingying and Mao, Bomin and Guo, Hongzhi and others},
  journal={IEEE Wireless Commun.},
  year={2026},
  month={Feb.},
  volume={},
  number={},
  pages={1-8},
  publisher={IEEE}
}

@article{mao2025blockchain,
  title={A blockchain-enabled cold start aggregation scheme for federated reinforcement learning-based task offloading in zero trust {LEO} satellite networks},
  author={Mao, Bomin and Liu, Yangbo and Wei, Zixiang and others},
  journal={IEEE J. Sel Areas Commun.},
  volume={43},
  number={6},
  pages={2172--2182},
  year={2025},
  month={Jun.},
  publisher={IEEE}
}

\section*{Biographies}

\vspace{-4mm}

\begin{IEEEbiographynophoto}{Xiangyu Li}
(xyli@eitech.edu.cn) 
received an M.S. degree from Georgia Institute of Technology, Atlanta, USA, in 2023, and he is currently pursuing a Ph.D. degree at Shanghai Jiao Tong University (SJTU), Shanghai, China, in the Eastern Institute of Technology (EIT)-SJTU Joint Ph.D. Program. His research interests include satellite communications, multiple access, and performance analysis of wireless systems.
\end{IEEEbiographynophoto}

\vspace{-4mm}

\begin{IEEEbiographynophoto}{Bodong Shang}
(bdshang@eitech.edu.cn) 
received his Ph.D. degree from Virginia Tech, Blacksburg, USA, in 2021, and he was a Postdoctoral Research Associate at Carnegie Mellon University, Pittsburgh, USA. 
He is currently an Assistant Professor at Eastern Institute of Technology (EIT), Ningbo, China. His research interests include space-air-ground-sea integrated networks, non-terrestrial networks, and space information networks.
\end{IEEEbiographynophoto}

\vspace{-4mm}

\begin{IEEEbiographynophoto}{Junchao Ma}
(junchao\_ma@jstu.edu.cn)
received the Ph.D. degree from Southwest Jiaotong University. He is currently a Lecturer with Jiangsu University of Technology. His current research interests include space information networks, Internet of Vehicles, and video caching.
\end{IEEEbiographynophoto}

\vspace{-4mm}

\begin{IEEEbiographynophoto}{Yuzheng Ren}
(renyuzheng@ustb.edu.cn) 
received the Ph.D. degree from Beijing University of Posts and Telecommunications, Beijing, China, in 2023. She is currently an Associate Professor at University of Science and Technology Beijing. Her research interests include intelligent networks and resource management.
\end{IEEEbiographynophoto}

\vspace{-4mm}

\begin{IEEEbiographynophoto}{Haijun Zhang}
(haijunzhang@ieee.org) 
is a Full Professor and Vice Dean at University of Science and Technology Beijing, China. He serves/served as an Editor of IEEE Transactions on Information Forensics and Security, IEEE Transactions on Communications, IEEE Transactions on Wireless Communications, and IEEE Transactions on Vehicular Technology. He is a Fellow of IEEE.
\end{IEEEbiographynophoto}

\vspace{-4mm}

\begin{IEEEbiographynophoto}{Pingyi Fan}
(fpy@tsinghua.edu.cn) 
is a Professor at the Department of Electronic Engineering of Tsinghua University and the Director of Open-Source Data Recognition Innovation Center. He is a member of the United States National Academy of Artificial Intelligence (US-NAAI), Co-chair of the academic committee of NAAI Asia Research Institute, and a Fellow of the IET.
\end{IEEEbiographynophoto}

\end{document}